%% file: iswcs_arxiv.tex
\documentclass[conference,11pt,onecolumn]{IEEEtran}

\usepackage{amsmath,dsfont,bbm,epsfig,amssymb,amsfonts,amstext,verbatim,amsopn,cite}
\usepackage{subfigure,multirow,multicol,lipsum,xfrac}
\usepackage{amsthm,ulem}
\usepackage{mathtools,amsthm}
\usepackage{perpage}
\usepackage{balance}
\usepackage{url}
\usepackage{amsfonts}
\usepackage{epsfig}
\usepackage[font={small}]{caption}
\usepackage{etoolbox}
\usepackage{algorithmicx}
\usepackage[Algorithm,ruled]{algorithm}
\usepackage{algpseudocode}
\usepackage{pifont}
\usepackage[utf8]{inputenc}
\usepackage[T1]{fontenc}  
\usepackage[nolist]{acronym}
\MakePerPage{footnote}
\usepackage{textcomp}
\usepackage{paralist}
\usepackage{enumitem}
\usepackage{bbm}
\usepackage[process=auto]{pstool}
\usepackage{tikz,pgfplots}
\usetikzlibrary{shapes,arrows}

\include{commands2}

\renewcommand{\emph}[1]{\textit{#1}}

\title{Multi-Antenna Towards Inband Shift Keying}
\author{
\IEEEauthorblockN{
Ralf R. M\"uller, \em Fellow, IEEE}
\IEEEauthorblockA{
Institute for Digital Communications, Friedrich-Alexander Universit\"at Erlangen-N\"urnberg, Germany\\
ralf.r.mueller@fau.de
}
}
\IEEEoverridecommandlockouts
\begin{document}
\renewcommand{\em}{\it}
\maketitle

\begin{abstract}
We propose a new continuous phase frequency shift keying that is particularly suited for multi-antenna communications when the link budget is critical and beam alignment is problematic. 
It combines the constant envelope of frequency modulation with low-rate repetition coding in order to compensate for the absence of transmit beamforming.
Although it is a frequency modulation, its transmit signal shows close to rectangular spectral shape. Similar to GSM's Gaussian minimum shift keying, it can be well approximated by linear modulation, when combined with differential precoding. This allows for easy coherent demodulation by means of a windowed fast Fourier transform.
\end{abstract}
\section{Introduction}
\label{Int}

Communication in the millimeter and terahertz bands are characterized by critical link budgets.
The standard solution to this problem is to utilize transmit beamforming which concentrates the signal energy around the receiving unit. However, this comes with lots of unfavorable issues:
1) Beam alignment delays the set-up of up radio links \cite{rangan:14}. 2) Directional transmission jeopardizes many forms of network coding. 3) The concentration of signal power raises concerns about electromagnetic compatibility \cite{silva:22}. 4) The narrow angular spread reduces multipath propagation \cite{durgin:00} and, thus, diversity. This list is not exhaustive. It only serves the purpose to motivate the search for an alternative to transmit beamforming.

Transmit beamforming is a spatial repetition code. On each antenna, the same data is transmitted.
If there are $N$ transmit antennas, the (spatial) code rate is $\frac 1N$.
In time domain, data symbols are conveniently separated by orthogonality via pulse shaping.
Coherent superposition in the vicinity of the receiver improves the signal-to-noise ratio (SNR) by a factor of $N^2$. 

Information theory does not care, if we exchange space and time on an additive white Gaussian noise (AWGN) channel. It just means to give the variables in the mathematical model of the channel another physical meaning. However, the hardware may be simpler to implement, if we apply the repetition code in time rather than in space. An the other hand, we may collect more noise, if we combine signals after receiver noise has been added.

Let all data be repeated $T$ times. The rate of this temporal repetition code is $\frac 1T$. 
In spatial domain, data symbols can be conveniently separated by orthogonality via frequency multiplex, if each subcarrier is transmitted over a different antenna. 
Coherent superposition within the receiver improves the useful signal by a factor of $T^2$, but the noise accumulates by a factor of $T$. So the SNR only improves by $T$, not $T^2$.

A multi-carrier multi-antenna system with temporal repetition coding that maps every subcarrier one to one onto the antenna elements is straightforward to implement with standard linear modulations.
However, this leaves much room for improvement. Linear modulation with pulse shaping also comes with one quite unfavorable issue: The amplitude varies in continuous time. The signal crest requires the amplifiers to be backed-off to avoid non-linear signal distortion and out-of-band radiation \cite{enzinger:18}. A significant amount of energy is converted into heat rather than being radiated.
Amplifiers which can cope with amplitude variations are also more expensive than their counterparts that can only cope with phase changes.

Nevertheless, linear modulation has displaced their nonlinear counterparts like continuous phase modulation and frequency shift keying over the evolution of wireless standard from 1G to 5G.
This was, as frequency shift keying comes with its own severe issues: 1) The spectrum is poorly confined. 2) The amplitude domain is not utilized for data transmission. Both effects result in poor spectral efficiency.

In a multi-carrier system with low-rate repetition coding, the negative issues of frequency shift keying, if properly designed, vanish, while the crest problems of linear modulation remain.
This makes continuous-phase frequency shift keying (CPFSK) a promising candidate for communication in millimeter and terahertz bands. The purpose of this paper is to propose one example of such a proper multi-carrier design of CPFSK, which is termed multi-antenna towards inband shift keying (MA-TISK), in the sequel.

For low-rate repetition coding, it is not an issue to leave the amplitude dimension of the signal unused and utilize the phase only. After all, a repetition code only utilizes one out of many dimensions in signal-space. On the AWGN channel, restricting to phase modulation has a very similar effect as a repetition code of rate one half, if the SNR is high. For low SNR, i.e.\ for critical link budget, the effect vanishes.

In order to exonerate CPFSK, at least for critical link budgets, it is sufficient to overcome its poor spectral confinement. A method to achieve that is presented in the following sections.
Section~\ref{RepCodGMSK} develops the main idea at the example of the frequency correction burst of the Global System for Mobile Communications (GSM).
Section~\ref{MA-TISK} generalizes the idea to arbitrary phase modulation and frequency pulses. Section~\ref{NumRes} provides numerical results, before Section~\ref{conc} summarizes the conclusions and promises.

\section{Repetition Coded GMSK}
\label{RepCodGMSK}

Data rate $R$ is fundamentally related to SNR.
For the complex-valued AWGN channel, the relation is well-investigated and given by
\begin{equation}
\label{shannon}
R = \log_2 (1+\mbox{SNR})
\end{equation}
If the link budget is critical, SNR is low, that means $\mbox{SNR}\ll 1$.
In that case, the logarithm in \eqref{shannon} is well approximated by a linear function.
We have
\begin{equation}
\label{shannon_approx}
R \approx \frac{\mbox{SNR}}{\ln 2}.
\end{equation}
In that regime, you can hardly do anything better than concatenating standard methods of forward-error correction coding designed for moderate rates with repetition coding in order to get an overall very low rate.

\begin{figure}
\centerline{\includegraphics[trim=40 5 40 5,clip,width=.9\columnwidth]{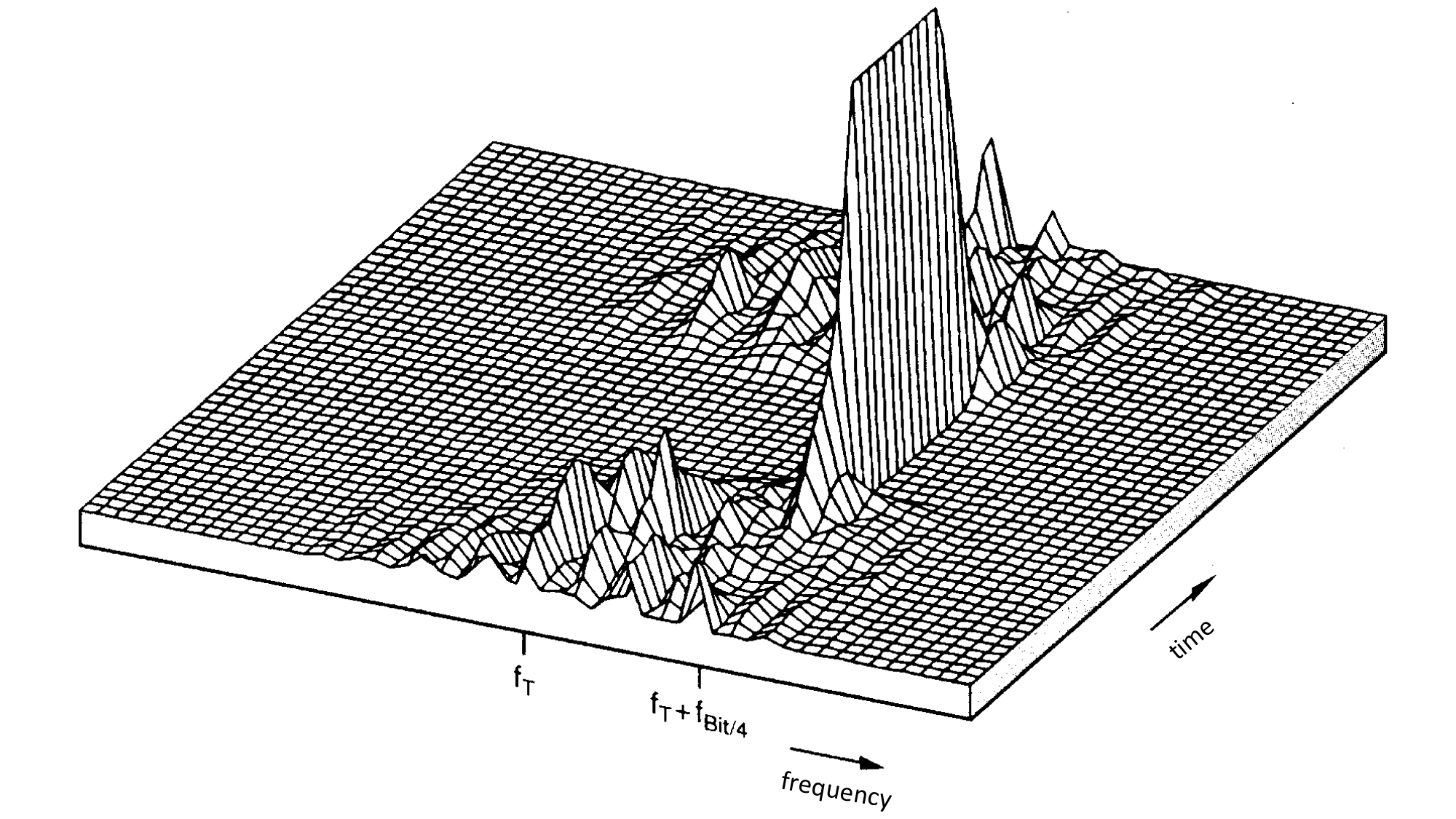}}
\caption{\label{FCB} Short-term spectrum during the frequency-correction burst of GSM \cite{frank:90}. In the figure, ${\rm f}_{\rm T}$ denotes the carrier frequency.}
\end{figure}

Let us consider repetition coding modulated with Gaussian minimum shift keying (GMSK) as used in GSM.
Consider a repetition code of rate $\frac1T$ with two codewords, the all-one and the all-zero word.
If we feed the all-zero codeword into the frequency modulator, the phase is repeatedly shifted upwards.
This is equivalent to an increase of frequency.
This effect is well-known and even utilized in GSM. Fig.~\ref{FCB} shows the short-term spectrum of GSM during the frequency correction burst which contains 148 consecutive 0-bits \cite{frank:90}.
This temporary sine tone at the shifted frequency is used in GSM for initial frequency synchronization and coarse time slot synchronization.

For the all-one codeword, the short-term spectrum looks very similar. The frequency shift is not towards lower frequencies, but also upwards. This is as GSM uses differential precoding. Thus, a sequence of 148 1-bits is first converted into a sequence of at least 147 0-symbols before it is fed into the modulator. For a duration of at least 142 symbols\footnote{GMSK modulation contains memory of up to 5 symbols.}, the instantaneous frequency is identical. However, the waveform is phase-shifted by $\pi$.

The repetition code composed of the all-zero and the all-one codeword shapes the spectrum of GMSK. The modified spectrum has a raised carrier frequency with reduced sidelobes.
During phase changes the sidelobes are predominantly left of the raised carrier frequency.
The spectrum is asymmetric.

An asymmetric spectrum is well-suited for a multi-carrier system. If the rightmost subcarrier had a spectrum as shown in Fig.~\ref{FCB}, the strong sidelobes to the left would interfere with lower subcarriers. Only the weak sidelobes to the right would contribute to out-of-band radiation. While this is good news for the top subcarrier, it is very bad at the lowest subcarrier frequency. There, the strong sidelobes would be out-of-band.

The spectrum in Fig.~\ref{FCB} can be flipped. For that purpose, we need to change the repetition code from ${\cal R}_{\rm u} = \{0000\dots,1111\dots\}$ to ${\cal R}_{\rm l}=\{1010\dots,0101\dots\}$. After differential precoding, this leads to an all-one sequence of at least 147 symbols. As MSK maps a logical $0$ to a phase shift of $\phi=\frac\pi2$ and a logical $1$ to a phase shift of $\phi=-\frac\pi2$, it results in a lowering of the instantaneous frequency.

The roadmap towards reduced out-of-band radiation should be clear now: We use the code ${\cal R}_{\rm u}$ for all subcarriers in the upper sideband (right of the center of the band) and the code ${\cal R}_{\rm l}$ for all subcarriers in the lower sideband (left of the center of the band).
Though one minor issue has remained: The frequencies of the subcarriers have shifted. Therefore, their orthogonality is lost. This, however, is easy to fix, as explained in the sequel.

Instead of changing the code, we could alternatively change the mapping.
Let us use the code ${\cal R}_{\rm u}$ for all subcarriers. However, for the subcarriers in the upper sideband, we use the mapping
\begin{equation}
\phi (s) = \begin{cases}
\frac \pi2 & \mbox{for} \qquad s = 0\\
-\frac \pi2 & \mbox{for} \qquad s = 1
\end{cases}.
\end{equation}
For subcarriers in the lower sideband, we use the mapping $ - \phi(s)$.
Clearly, this gives the same result, as the two codes just differ in the sign of the symbols fed into the modulator.

We can correct the frequency shift by an offset to the phase shifts, as instantaneous frequency is the derivative of the phase. This means, we use the mappings
\begin{align}
\phi_{\rm u}(s) &= \phi(s)  -\frac\pi2 = \begin{cases}
0 & \mbox{for} \qquad s = 0\\
- \pi & \mbox{for} \qquad s = 1
\end{cases}\\
\phi_{\rm l}(s) &=  \frac\pi2 -\phi(s)
= \begin{cases}
0 & \mbox{for} \qquad s = 0\\
\pi & \mbox{for} \qquad s = 1
\end{cases}
\end{align}
for the upper and the lower sideband, respectively.
Here, it is important to note that a phase shift by $\pi$ and by $-\pi$ are not identical in CPFSK.
A positive phase shift means that the instantaneous frequency is temporarily increased, while a negative phase shift means that it is temporarily reduced.
After the decay of the frequency pulse, the result does not differ, of course.
If the code rate $\frac1T$ is small, the receiver will hardly notice the difference, either, but the radio spectrum is strongly affected.

\subsection{A Misconception of GSM}
GSM is standardized with minimum shift keying as modulation format.
This was based on the thought that minimum shift keying is the most efficient frequency modulation, which is correct \cite{proakis:95}.
However, GMSK is combined with differential precoding. Thus, it is a differential frequency modulation. In that case, minimum shift keying is not proven optimal. For the time-bandwidth product of $BT=0.3$ in GSM, it can be well-approximated by linear modulation \cite{mouly:92}. In this case, it is certainly not optimal.

\begin{figure}
\centerline{\includegraphics[trim=10 1 50 20,clip,width=.70\columnwidth]{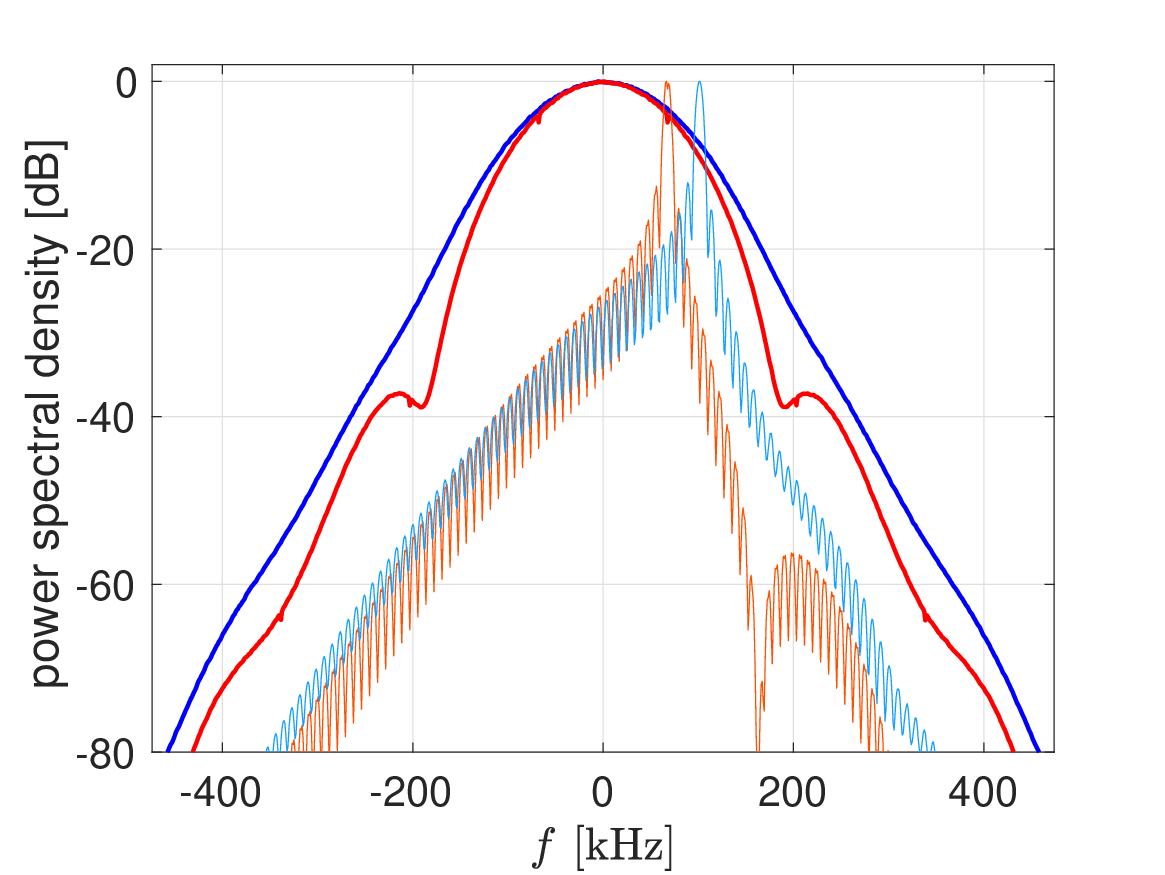}}
\caption{\label{GSMspectrum} Spectrum of GSM in bold red line. Spectrum for quaternary GMSK in bold blue line. For comparison the asymmetric spectra with repetition codes of rate $\frac1{32}$ are shown in thin lines repeating phase shifts of $\frac\pi2$ and $\frac{3\pi}4$, respectively.}
\end{figure}

Consider a quaternary minimum shift keying \cite{ekanayake:82} with Gaussian pulse filtering, i.e.\  a generalization of GMSK where everything stays as it is, except for having quaternary phase-shift keying (QPSK) symbols instead of its binary counterpart.
Instead of the phase shifts $\pm \frac\pi2$, we have phase shifts from the set $\{\pm\frac\pi4,\pm\frac{3\pi}4\}$ and apply Gray mapping.
Both quadrature components can be demodulated independently from each other \cite{ekanayake:82}.
For the QPSK example above, the maximal frequency deviation is $\frac{3\pi}4$ instead of $\frac\pi2$. This results in an at most 37 \% wider (slightly above the shoulder of the first sidelobe) spectrum as shown in Fig.~\ref{GSMspectrum}.
However, it doubles data rate. This is a boost in spectral efficiency of at least 46\% that was not realized. An alternative way to utilize the inherent redundancy of binary GMSK is reported in \cite{meyer:06}.

In our multi-antenna concept, we are only concerned with the spectra for repetition codes. As Fig.~\ref{GSMspectrum} illustrates, the spectrum of quaternary GMSK is hardly worse than the spectrum of its binary counterpart. Note that only the spectrum of the upper sideband relative to the main lobe influences the out-of-band radiation. The spectrum of the lower sideband only contributes to inter-carrier interference.

\subsection{Asymmetric Quaternary GMSK}

\begin{figure}
\centerline{\includegraphics[trim=10 1 30 20,clip,width=.7\columnwidth]{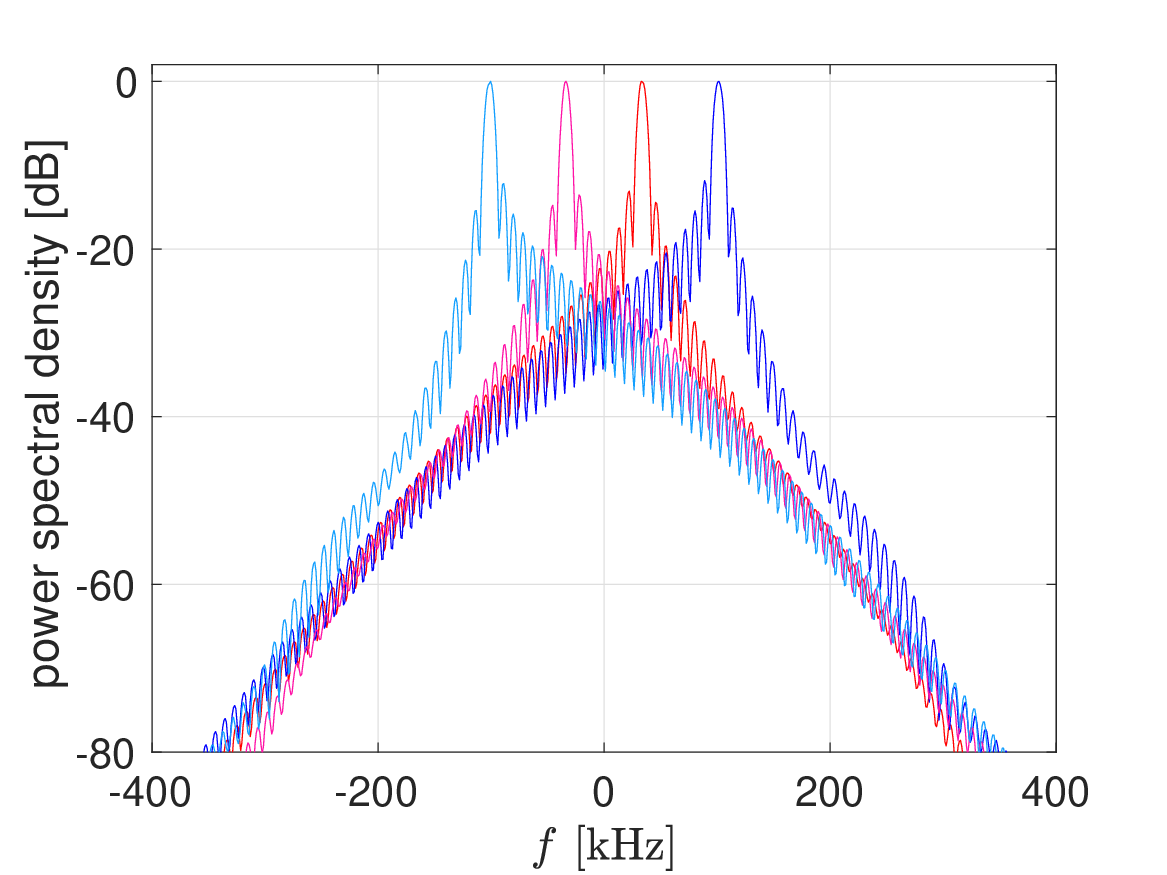}}
\caption{\label{QGMSK} Spectra of quaternary GMSK with repetition coding of rate $\frac1{32}$ and phase constellation $\{\pm \frac\pi2, \pm \frac{3\pi}2\}$.
In the leftmost and rightmost curve, the phase shifts $-\frac{3\pi}2$ and $\frac{3\pi}2$ are repeated, respectively. Same for the inner curves and phase shifts of $\pm\frac\pi2$.}
\end{figure}

In contrast to binary GMSK discussed in Section~\ref{RepCodGMSK}, a modification of the phase mapping is not just a frequency shift and/or mirroring, when a repetition code is used. For  quaternary phase constellations, it makes a difference, whether the repeated phase shift, i.e.\ the one that the symbol $s=0$ is mapped to, is at the edge of the constellation, i.e.\ has minimum or maximum phase shift, or it is an inner phase shift between these extremes.
As noted before, a phase shift by a multiple of $2\pi$ is not irrelevant for the spectrum of CPFSK.
The four different spectra are shown in Fig.~\ref{QGMSK}.
There are two different shapes with their respective mirrored spectra.

\begin{figure}
\centerline{\includegraphics[trim=10 1 50 20,clip,width=.7\columnwidth]{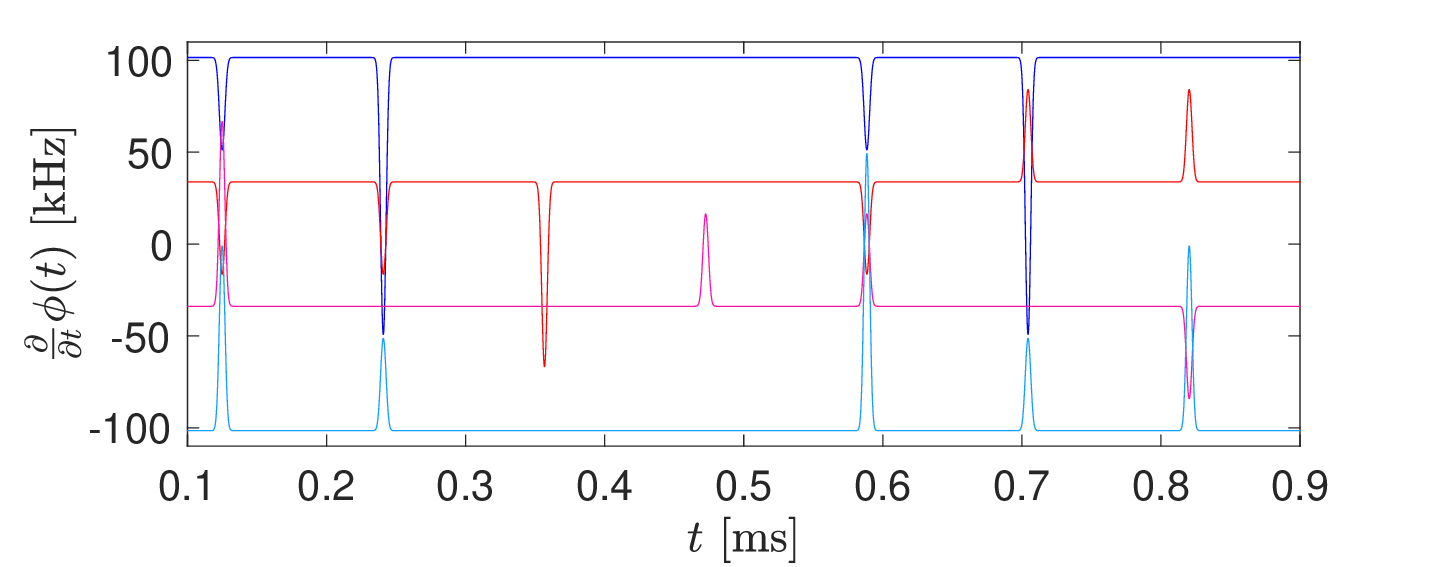}}
\caption{\label{instfreqQGMSK} Instantaneous frequencies of quaternary GMSK with repetition coding of rate $\frac1{32}$ and random QPSK data.
In the lower and upper curve, the phase shifts $-\frac{3\pi}2$ and $\frac{3\pi}2$ are repeated, respectively. Same for the inner curves and phase shifts of $\pm\frac\pi2$.}
\end{figure}

The repeated phase shifts determine the instantaneous frequency.
This can be observed in Fig.~\ref{instfreqQGMSK}.

For a multi-carrier system, spectra as in Fig.~\ref{QGMSK} are quite useful. Each could represent one subcarrier.
The overall spectrum of all tones is still better confined as its counterpart without repetition coding in Fig.~\ref{GSMspectrum}.
However, the carrier spacing in Fig.~\ref{QGMSK} is quite wasteful. Many more tones could be packed in-between these four ones. This issue is addressed in the next section.
\section{Multi-Antenna Towards Inband Shift Keying}
\label{MA-TISK}

Differential precoding converts any constellation of phases into a constellation of phase shifts.
Any regular PSK constellation with $M$ points is mapped onto the set of phase shifts
\begin{equation}
{\cal P} = \left\{0, \frac{2\pi}M, \dots, \frac{2\pi(M-1)}M\right\}.  
\end{equation}
While a symbol is repeated, no phase shift occurs and the instantaneous frequency keeps constant.

\subsection{Antenna/Subcarrier-Dependent Constellation}
We can optimize the spectral shape without influencing the performance, e.g., error and information rate, of a digital communication system.
In discrete time, a phase shift by $\phi$ and by $\phi+2\pi$ cannot be distinguished.
In continuous time, however, the output signals of a CPFSK modulator differ. For the latter case, the instantaneous frequency is larger than for the former.
This influences the spectral shape.

Consider the general set of PSK phase shifts
\begin{equation}
{\cal P(\bomega)} = \left\{0, \frac{2\pi}M-2\pi \omega_1, \dots, \frac{2\pi(M-1)}M-2\pi\omega_{M-1}\right\}.  
\end{equation}
for some binary vector $\bomega=[\omega_1,\dots,\omega_{M-1}]\in\{0,1\}^{M-1}$.
For each subcarrier, we can find the optimal vector $\bomega$, such that the spectrum of the aggregated signal of all subcarriers is most confined. 

For the leftmost subcarrier the optimal choice is the all-zero vector, if the number of subcarriers is large.
For the rightmost subcarrier, it is the all one-vector.
This is, as we prefer the sidelobes to appear towards the center of the channel band.

For other subcarriers, the optimization can be tedious and requires a clear way to measure spectral confinement. This task is left open for future work.
In this work, we opt for an heuristic, but reasonable approach: We chose $\bomega$ for each subcarrier in such a way that the maximal deviation of the instantaneous frequency from the center of band is as small as possible.
We call this method, as well as its generalizations with more sophisticated optimization of the vector $\bomega$, multi-antenna towards inband shift keying.
\subsection{Pulse Waveforms}

In GSM, a Gaussian filtered rectangular frequency pulse shape is used. It is well optimized with respect to the trade-off between spectral confinement and limiting intersymbol interference caused by lack of temporal orthogonality.
In MA-TISK, the particular choice of the frequency pulse shape is much less critical.

The repetition code ensures that the shape of the pulse hardly causes intersymbol interference. Due to the repetition code, the transmit signal is similar to a direct-sequence spread-spectrum signal with a row of the discrete Fourier transform matrix as spreading sequence.
The frequency pulse shape corresponds to the chip pulse in the spread-spectrum signal. Even chip pulses that are not shift-orthogonal hardly create intersymbol interference. They may interfere with few subsequent chips. But they hardly span long enough in time to cause relevant intersymbol interference.
The same considerations apply to the frequency pulses in MA-TISK.

The design of frequency pulses can overwhelmingly focus on spectral confinement. As argued above, intersymbol interference is not a critical issue due to the repetition coding. For GMSK pulses, we can thus go for lower time-bandwidth products than in GSM. I
Optimization of the frequency pulse shape is left open for future work.

\subsection{Subcarrier Separation}

Ideally, the subcarriers are spaced in such a way that they can be easily demodulated via fast Fourier transform (FFT) \cite{cooley:65} while keeping the inter-carrier interference at a minimum.

For $N$ subcarriers, the rate $R$ of the repetition code should be slightly smaller than $\frac1N$.
If we repeat for $T>N$ times, the frequency pulse has time to decay such that the instantaneous frequency is constant for $N$ symbol periods.
In that case, we can sample the received signal, window it for $N$ symbol periods and perfectly separate the $N$ subcarriers via FFT. In practice, some crosstalk may remain. The frequency pulse may not have fully decayed or channel dispersion may jeopardize the orthogonality.  These effects can be counteracted by slightly reducing the rate $R$, as demonstrated in Section~\ref{NumRes}.

The system may even work with severe crosstalk.
In \cite{mueller:11,mueller:11a} a system was proposed that uses non-orthogonal spreading sequences with constant envelope instead of subcarriers. With the help of multiuser detection \cite{mueller:11b,gallinaro:17}, it is successfully utilized in \cite{itu-rm.2092-1:22} for satellite communications with a tight spectral mask that asks for a stopband attenuation of $-70$~dB.


\section{Numerical Results}
\label{NumRes}

We show an example to demonstrate the quantitative potential of MA-TISK. We compare the resulting spectrum to various single-carrier methods that are state-of-the-art.

We target the spectral mask of 5G downlink at 100 MHz channel bandwidth. We choose a transmit array with $N=64$ antennas/subcarriers and QPSK modulation at a symbol rate of 1.25 Msymbols/s on each subcarrier resulting in a total data rate of 160 Mbit/s. 
The corresponding symbol period is 800 ns. 
We take a Gaussian filtered rectangular pulse like in the GSM standard \cite{mouly:92} but with time-bandwidth product of 0.1 and scale it in time to fit the new data rate. This leads to a pulse duration of 91.4 ns. 

In order to have the GMSK pulse largely decayed before the FFT window starts, we set $T=70$. This results in a subcarrier spacing of 1.37 MHz.
This is, by far, wide enough to be insensitive to Doppler-induced frequency shifts.

\begin{figure}
\centerline{\includegraphics[trim=1 1 20 40,clip,width=.6\columnwidth]{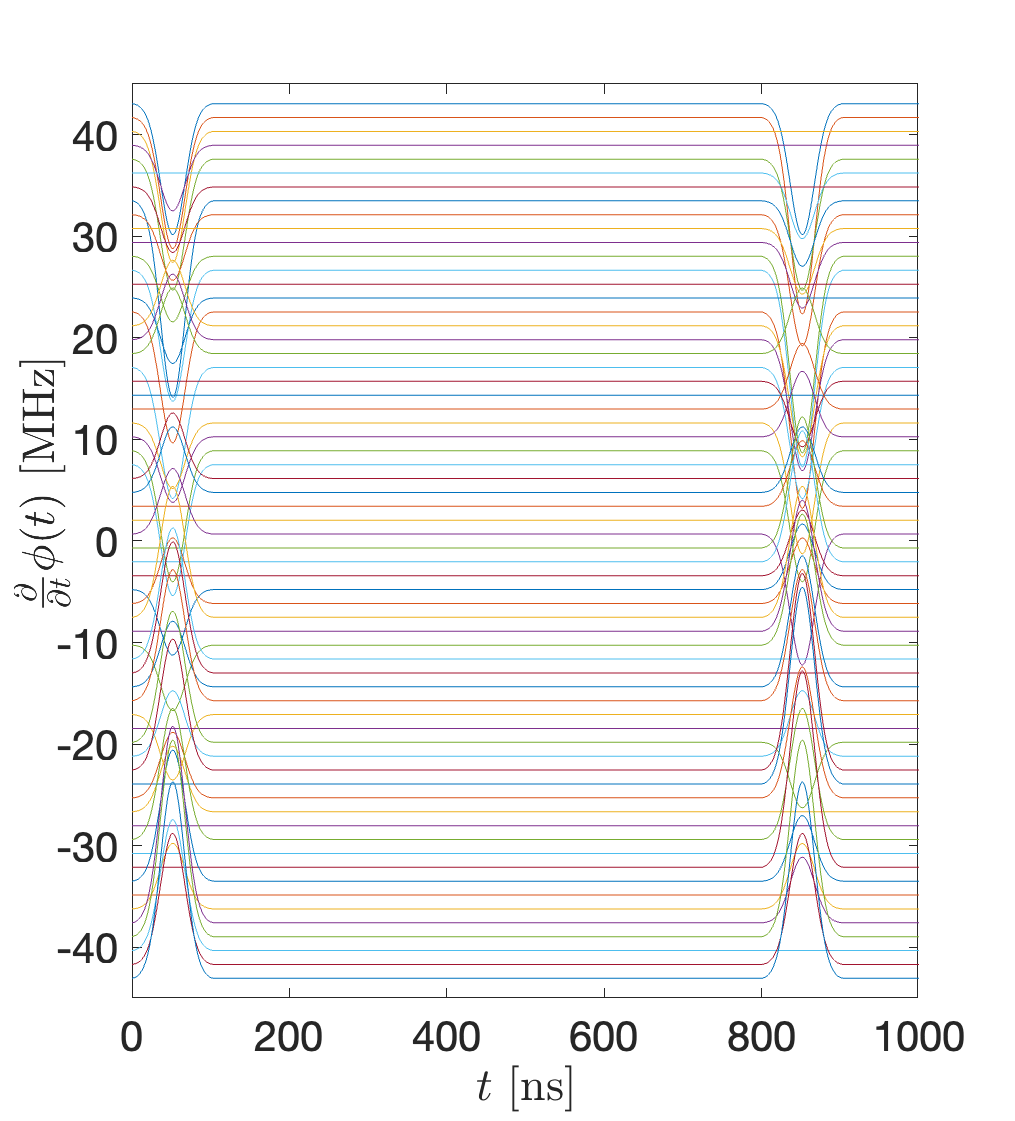}}
\caption{\label{instfreq64}
Instantaneous frequencies of 64 subcarriers with Gaussian filtered rectangular pulses.
}
\end{figure}

The resulting instantaneous frequencies are shown in Fig.~\ref{instfreq64} for 1.25 symbol periods and all 64 subcarriers.
The instantaneous frequencies change over time, but they always keep between the lowest and highest subcarrier frequency. Changes are predominantly directed towards the center of the frequency band.

For most of the symbol period, MA-TISK is linear phase modulation. The instantaneous frequency is constant.
The phase of each subcarrier corresponds to the PSK symbol to be transmitted. Only during phase transitions, when the instantaneous frequency changes, the modulated signal relates to the PSK symbol in a nonlinear way. For many antennas, this period is much shorter than the symbol period and can be dismissed in a similar way as the cyclic prefix in orthogonal frequency division multiplexing.

The more antennas are utilized, the larger the symbol period, as the overall bandwidth is divided into more subbands. This makes MA-TISK the more robust against channel dispersion the more antennas are used.

\begin{figure}
\centerline{\includegraphics[trim=10 1 50 19,clip,width=.7\columnwidth]{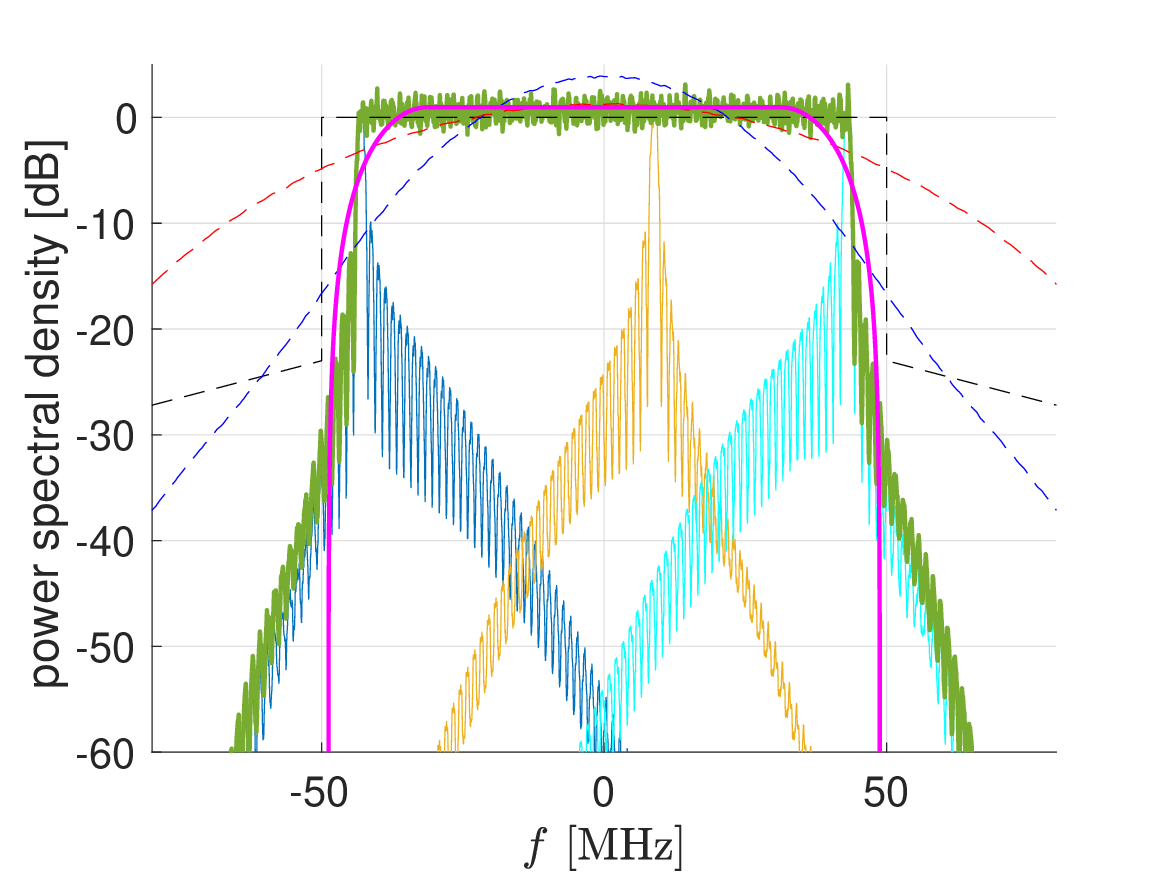}}
\caption{\label{comPSD}
PSD of the transmit signal vs.\ frequency for total data rate of 160 Mbit/s.
The green line shows the average spectrum.
The thin lines in blue, yellow, and cyan refer to the spectra of the leftmost, an intermediate, and the rightmost subcarrier, respectively.
The red and blue dashed lines refer to the spectra of transmit beamforming when using binary and quaternary GMSK modulation. The pink curve refers to transmit beamforming with linear QPSK and a root-raised cosine pulse with roll-off factor 0.22. The black dashed line is the spectral mask of 5G. PSDs are normalized to have the same total power.}
\end{figure}

The power spectral density (PSD) of the transmit signal is shown in Fig.~\ref{comPSD}.
While the aggregate signal is wideband, the transmit signals at any individual antenna element is very narrowband.
This simplifies the design of the RF-circuitry.
In fact, narrowband RF-circuitry would even help to suppress sidelobes and reduce inter-carrier interference.
Furthermore, the asymmetry of the PSD at the leftmost and rightmost subcarriers is evident.
Even the spectrum of the centered carrier is not symmetric, as it performs phase shifts by $-\pi$, but not by $+\pi$.
The overall spectrum shows significantly less sidelobes than the spectra of the individual subcarriers.
This is a direct result of the in-band direction of the frequency deviations.

When compared to single-carrier GMSK, the spectrum of MA-TISK is far better confined and competitive with linear single-carrier PSK. The latter has slight advantages when it comes to high stopband attenuation, but its spectrum deviates more from the ideal rectangular shape in the utilized spectral range. 

After demodulation by means of sampling and FFT,
the signal-to-interference ratio (SIR) is found as 34.4 dB. No Gaussian noise was added. Interference is solely due to crosstalk of subcarriers and intersymbol interference.
The SIR heavily depends on the value of $T$, cf.\ Table~\ref{Ttab},
\begin{table}
\caption{\label{Ttab}SIR for various numbers of repetitions $T$ given a fixed number $N=64$ of antennas/subcarriers.}
\begin{center}
\begin{tabular}{ccccccccccccccc}
\hline
$T\vphantom{\frac12}$ &64 & 65 & 66 & 67 & 68 & 69 & 70 & 71 & 72 & 73\\
\hline
\!\!\!\!SIR [dB]$\vphantom{\frac12}$& 8.1 & 9.5 & 11 &13 & 17 & 24 & 34 & 49 & 72 &110\\
\hline
\end{tabular}
\end{center}
\end{table}
and the duration of the frequency pulse.
For $T=64=N$, the frequency pulse is fully included in the FFT. This results in significant inter-carrier interference.
Though, in combination with strong forward error-correction coding, an SIR of 8~dB is more than sufficient for reliable communication with quaternary modulation.
For increasing values of $T$, the window of the FFT begins to span only that range of the symbol periods where the instantaneous frequency is constant.
Thus, the subcarriers preserve orthogonality.
 
The temporal combining gain is 18.1~dB. Note that the temporal combining gain can slightly exceed 18~dB as $T=70 >N=64$.
Setting $T>N$ is spectrally wasteful. However, it reduces inter-carrier interference, cf.\ Table~\ref{Ttab}. Note that linear single-carrier modulation is also spectrally wasteful due to the roll-off factor.

\section{Conclusions \& Promises}
\label{conc}

Link budget can be boosted not only by transmit beamforming, but also by temporal repetition coding, although less efficiently from a pure link budget perspective.
However, the lower boost in link budget may be compensated for by considerations on link set-up and hardware costs.

Temporal repetition coding is conveniently implemented via a multi-carrier system where each subcarrier is mapped onto one transmit antenna and differentially precoded continuous phase frequency shift keying avoids the spectral issues that come with frequency modulation in single-carrier systems. MA-TISK, a promising method of such implementation, is proposed.

MA-TISK utilizes sets of phase shifts that depend on the position of the subcarrier within the channel band. This tilts the spectrum of each subcarrier towards the center of the band and reduces sidelobes towards adjacent channels.

Although MA-TISK is a non-linear modulation, it can be demodulated like a linear one thanks to its combination with repetition coding and differential precoding.
Intersymbol interference due to the memory of the continuous phase modulation is negligible.
Cost-efficient demodulation by means of fast Fourier transform is possible.

The transition phase may introduce some minor overhead. However, the overheads of competing systems, e.g., roll-off factors or blank subcarriers, are avoided.

Thanks to its constant envelope, MA-TISK can be efficiently implemented by voltage controlled oscillators.
Power amplifiers need neither be linear nor broadband and can operate without back-off at high efficiency.

MA-TISK does not require beam alignment. Radio links can be set up fast without serious overhead. This makes MA-TISK suitable for the implementation of internet of things (IoT)-applications in the millimeter-wave and terahertz bands.

MA-TISK is the more sensitive to multipath propagation the fewer antennas are deployed. This makes MA-TISK the more competitive the higher the carrier frequency. 

\section*{Acknowledgement}
The author would like to thank M.\ Vossiek, C.\ Carlowitz, \& L.\ Hahn for discussions on millimeter wave circuitry. Further thanks go to G.\ Fischer, W.\ Gerstacker, B.\ Niemann, A.\ Mezghani, and H.\ Rosenberger for discussions on RF amplifiers, frequency domain equalization, 6G, and transmit beamforming, as well as proof reading the manuscript, resp.
\bibliography{lit}
\bibliographystyle{unsrt}
\end{document}

%% file: commands2.tex
\newcommand{\bomega}{{\boldsymbol{\omega}}}

\newtheoremstyle{mystyle}
  {}
  {}
  {}
  {}
  {\bfseries}
  {:}
  { }
  {}
\theoremstyle{mystyle}

\algnewcommand\algorithmicLet{\textbf{Let}}
\algnewcommand\Let{\item[\algorithmicLet]}
\algnewcommand\algorithmicSet{\textbf{Set}}
\algnewcommand\Set{\item[\algorithmicSet]}

\algnewcommand\algorithmicInitiate{\textbf{Initiate}}
\algnewcommand\Initiate{\item[\algorithmicInitiate]}
\algnewcommand\algorithmicStart{\textbf{Begin}}
\algnewcommand\Begin{\item[\algorithmicStart]}
\algnewcommand\algorithmicEnd{\textbf{End}}
\algnewcommand\End{\item[\algorithmicEnd]}

\algnewcommand\algorithmicOutP{\textbf{Output:}}
\algnewcommand\Out{\item[\algorithmicOutP]}

\algnewcommand\algorithmicInP{\textbf{Input:}}
\algnewcommand\In{\item[\algorithmicInP]}

\newcounter{bar}

